\documentstyle[epsf,eqsecnum,floats,preprint,aps]{revtex}

\tighten

\input epsf
\begin{document}

\newcommand{\beq}{\begin{equation}}
\newcommand{\eeq}{\end{equation}}

\newcommand{\be}{\begin{equation}}
\newcommand{\ee}{\end{equation}}
\newcommand{\bea}{\begin{eqnarray}}
\newcommand{\eea}{\end{eqnarray}}
\newcommand{\PSbox}[3]{\mbox{\rule{0in}{#3}\includegraphics{#1}\hspace{#2}}}

\def\5M{M^3_{(5)}}
\def\4M{M^2_{(4)}}

\def\d{\delta}
\def\fourG{{{}^{(4)}G}}
\def\H{{\cal H}}
\def\K5{{\kappa}}
\def\K52{{\kappa^2}}
\def\C{{\cal C}}
\def\lamb{{\rho_{\Lambda}}}

\newcommand{\A}{A}
\newcommand{\B}{B}
\newcommand{\mmu}{\mu}
\newcommand{\mnu}{\nu}
\newcommand{\ii}{i}
\newcommand{\jj}{j}
\newcommand{\jl}{[}
\newcommand{\jr}{]}
\newcommand{\ml}{\sharp}
\newcommand{\mr}{\sharp}

\newcommand{\da}{\dot{a}}
\newcommand{\db}{\dot{b}}
\newcommand{\dn}{\dot{n}}
\newcommand{\dda}{\ddot{a}}
\newcommand{\ddb}{\ddot{b}}
\newcommand{\ddn}{\ddot{n}}
\newcommand{\pa}{a^{\prime}}
\newcommand{\pb}{b^{\prime}}
\newcommand{\pn}{n^{\prime}}
\newcommand{\ppa}{a^{\prime \prime}}
\newcommand{\ppb}{b^{\prime \prime}}
\newcommand{\ppn}{n^{\prime \prime}}
\newcommand{\fda}{\frac{\da}{a}}
\newcommand{\fdb}{\frac{\db}{b}}
\newcommand{\fdn}{\frac{\dn}{n}}
\newcommand{\fdda}{\frac{\dda}{a}}
\newcommand{\fddb}{\frac{\ddb}{b}}
\newcommand{\fddn}{\frac{\ddn}{n}}
\newcommand{\fpa}{\frac{\pa}{a}}
\newcommand{\fpb}{\frac{\pb}{b}}
\newcommand{\fpn}{\frac{\pn}{n}}
\newcommand{\fppa}{\frac{\ppa}{a}}
\newcommand{\fppb}{\frac{\ppb}{b}}
\newcommand{\fppn}{\frac{\ppn}{n}}

\newcommand{\dA}{\dot{A_0}}
\newcommand{\dB}{\dot{B_0}}
\newcommand{\fdA}{\frac{\dA}{A_0}}
\newcommand{\fdB}{\frac{\dB}{B_0}}

\overfullrule=0pt
\def\Int{\int_{r_H}^\infty}
\def\d{\partial}
\def\e{\epsilon}
\def\M{{\cal M}}
\def\high{\vphantom{\Biggl(}\displaystyle}
\catcode`@=11
\def\@versim#1#2{\lower.7\p@\vbox{\baselineskip\z@skip\lineskip-.5\p@
    \ialign{$\m@th#1\hfil##\hfil$\crcr#2\crcr\sim\crcr}}}
\def\simge{\mathrel{\mathpalette\@versim>}} %
\def\simle{\mathrel{\mathpalette\@versim<}} %
\catcode`@=12 

\def\pr#1#2#3#4{Phys. Rev. D {\bf #1}, #2 (19#3#4)}
\def\prl#1#2#3#4{Phys. Rev. Lett. {\bf #1}, #2 (19#3#4)}
\def\prold#1#2#3#4{Phys. Rev. {\bf #1}, #2 (19#3#4)}
\def\np#1#2#3#4{Nucl. Phys. {\bf B#1}, #2 (19#3#4)}
\def\pl#1#2#3#4{Phys. Lett. {\bf #1B}, #2 (19#3#4)}

\rightline{NYU-TH 01/01/03}
\rightline{TPI-MINN-01/06}
\rightline{UMN-TH-1937}
\rightline{hep-th/0104201}
\vskip 1cm

\setcounter{footnote}{0}

\begin{center}
\Large{\bf Braneworld Flattening by a Cosmological Constant}
\ \\
\ \\
\normalsize{Cedric Deffayet$^\dagger$\footnotetext{deffayet@physics.nyu.edu},
Gia Dvali$^\dagger$\footnotetext{dvali@physics.nyu.edu},
Gregory Gabadadze$^\ddagger$\footnotetext{gabadadz@physics.umn.edu},
and Arthur Lue$^\dagger$\footnotetext{lue@physics.nyu.edu}}
\ \\
\ \\
$^\dagger$\small{\em Department of Physics \\
New York University \\
New York, NY 10003}
\ \\
\ \\
$^\ddagger$\small{\em Theoretical Physics Institute \\
University of Minnesota\\
Minneapolis, MN 55455}

\end{center}

\begin{abstract}

\noindent
We present a model with an infinite volume bulk in which a braneworld
with a cosmological constant evolves to a static, 4-dimensional
Minkowski spacetime.  This evolution occurs for a generic class of
initial conditions with positive energy densities.  The metric
everywhere outside the brane is that of a 5-dimensional Minkowski
spacetime, where the effect of the brane is the creation of a frame
with a varying speed of light.  This fact is encoded in the structure
of the 4-dimensional graviton propagator on the braneworld, which may
lead to some interesting Lorentz symmetry violating effects.  In our
framework the cosmological constant problem takes a different meaning
since the flatness of the Universe is guaranteed for an arbitrary
negative cosmological constant.  Instead constraints on the model come
from different concerns which we discuss in detail.
\end{abstract}

\setcounter{page}{0}
\thispagestyle{empty}
\maketitle

\eject

\vfill

\baselineskip 18pt plus 2pt minus 2pt

\section{Introduction}

\setcounter{footnote}{0}

Understanding why the vacuum energy density is essentially zero is a
fundamental challenge of contemporary physics.  The vacuum energy
density is a free parameter of nature unprotected from large quantum
corrections and, therefore, an exquisitely small value for this
parameter is an unexplained fine-tuning.

One approach to dealing with this problem \cite{Dvali:2000rv,witten} relies on
braneworld theories with {\it infinite volume} extra dimensions.
The first attempt of such models appeared in \cite{quasi1};  however,
we consider theories based on a different approach
\cite{Dvali:2000hr,Dvali:2000xg,Deffayet,Dvali:2001gm}.
These theories exhibit the following unique properties.  First, gravity
becomes higher-dimensional at large distances; and second, the
infinite volume extra dimensions allow for exact bulk supersymmetry
(compatible with SUSY broken on the brane) which can control the value
of the bulk cosmological term.
As a result, it is natural to expect that the large scale cosmology is
that of the higher-dimensional theory with a cosmological term that
can be naturally zero due to unbroken bulk SUSY.  Even so, there
are questions that must be addressed before such a solution
can be regarded as viable:

\begin{itemize}
\item
What is the mechanism that makes gravity appear 4-dimensional
on the brane?
\item
Can the Universe evolve to the desired (present) state without
sacrificing successful features of standard
Friedmann--Lema{\^{\i}}tre--Robertson--Walker (FLRW) cosmology
during the earlier stages of the evolution?
\end{itemize}

To provide an answer to the first question we work in the
framework of models with an induced intrinsic curvature term on the
brane \cite{Dvali:2000hr,Dvali:2000xg,Deffayet,Dvali:2001gm}.  In this
framework gravity is 4-dimensional up to astronomical distances
due to the large graviton kinetic term on the brane.  Explicit
cosmological solutions confirming large distance crossover behavior
within this class of theories were already found in\cite{Deffayet}.

In the present paper we discuss a particular class of those
cosmological scenarios where the cosmological constant on the brane is
negative.  Irrespective of the magnitude of this constant, the
4-dimensional metric on the brane {\it automatically} evolves to a
static Minkowski spacetime for generic initial conditions with
positive energy density.  This attractive scenario does not requires
fine-tuning; however, it cannot be regarded as a solution of the
cosmological constant problem since phenomenological considerations
require a small brane cosmological constant.  Nevertheless, the
example is rather remarkable since the meaning of cosmological
constant problem is changed.  Unlike the usual 4-dimensional scenario
where the smallness of the vacuum energy is required for the observed
flatness of the Universe, in the present framework the Universe is
automatically flat for arbitrarily large negative vacuum energy.
The constraint on the brane cosmological constant comes from
completely different considerations, such as ultralarge distance
gravity measurements and 4-dimensional FLRW--cosmological history.
In other words, in our scenario we ask not: ``Why is the Universe
flat?'' but rather: ``Why does the Universe have FLRW--type history?''

Finally let us note that a byproduct of our scenario is small
Lorentz-violating effects,\footnote{ The metric on the brane is
Minkowskian, even though the energy momentum tensor is nonzero and
does not respect Lorentz symmetry.  The metric of the 5-dimensional
space, which is supported solely by the energy-momentum on the
braneworld, also does not respect 4-dimensional Lorentz symmetry. Any
slice of the 5-dimensional spacetime parallel to the braneworld is
indeed Minkowski, however, the speed of light varies from slice to
slice.  Gravity can probe between slices and is capable of explicitly
reflecting this 4-dimensional Lorentz symmetry violation.} for which
there is a growing interest in both particle phenomenology contexts
(see for example \cite{Lorentz1,Lorentz2}), as well as in braneworld
scenarios \cite{Visser:1985qm,DvaliShif,CEG,Dubovsky:2001fj}.

We first discuss the ingredients necessary to construct our model, as
well as describe its cosmological evolution and the global
5-dimensional structure of spacetime.  We then go on to evaluate the
propagation of perturbations on the braneworld itself and
phenomenological constraints. We discuss these limitations in detail
in the concluding remarks.

\section{The Solution, Dynamics and Cosmology}

Consider a three-brane embedded in a 5-dimensional spacetime.  The
bulk is empty; all energy-momentum is isolated on the brane.  The
action is
\beq
S_{(5)}=-\frac{1}{2}M^3_{(5)} \int d^5x
\sqrt{-\tilde{g}} \tilde{R} 
+\int d^4x \sqrt{-g}{\cal L}_m \ .
\label{action} 
\eeq
The first term in Eq.~(\ref{action}) corresponds to the Einstein-Hilbert
action in five dimensions for a 5-dimensional metric ${\tilde
  g}_{AB}$ (bulk metric) with Ricci scalar $\tilde{R}$.  This action
is sufficient to generate our flattening cosmological solution.
However, in order to obtain 4-dimensional gravitation on the braneworld
as well as 4-dimensional cosmology, we also consider
an intrinsic curvature term which is generally induced by radiative
corrections by the matter density on the brane \cite{Dvali:2000hr}:
\beq
-\frac{1}{2}M^2_{(4)} \int d^4x
\sqrt{-{g}} {R}\ .
\label{branac}
\eeq
Similarly, Eq.~(\ref{branac}) is the Einstein-Hilbert action for the
induced metric
$g_{cd}$ on the brane, $R$ being its scalar curvature. The induced
metric\footnote{Throughout this article, we adopt the following
convention for indices: upper case Latin letters $A,B,...$
  denote 5D indices: $0,1,2,3,5$; lower case Latin letters from the
  beginning of the alphabet: $c,d,...$ denote 4-dimensional
   indices parallel
  to the brane, lower case Latin letters from the middle of the
  alphabet: $i,j,...$ denote space-like 3D indices parallel to
  the brane.}  $g_{cd}$ is defined as usual from the bulk metric
${\tilde g}_{AB}$ by
\beq \label{induite}
g_{cd} = \partial_c X^A \partial_d X^B {\tilde g}_{AB}\ ,
\eeq
where $X^A(x^c)$ represents the coordinates of an event on the brane
labeled by $x^c$.  We neglect higher-derivative terms in the bulk
and worldvolume actions as they lead to the modification of gravity
at undetectably short distances.

For a brane embedded in a Minkowski spacetime with such an action, it
is shown in Ref.~\cite{Dvali:2000hr} that the usual 4-dimensional
Newton's law for static point like sources on the brane is recovered
at observable distances.  At ultralarge cosmological
distances the gravitational force is given by the 5-dimensional
$1/r^3$ force law.  The crossover length scale between the two
different regimes is given by
\beq \label{crosso}
r_0 =\frac{M_{(4)}^2}{2M_{(5)}^3}\ . 
\eeq
We consider a system where there is a negative cosmological constant
as well as some arbitrary matter localized on the brane.
Another mass scale that arises in our discussion is
\beq
\mu^2 = {-\lambda \over M^2_{(4)}}\ ,
\eeq
where $\lambda$ is the (negative) cosmological constant on the brane.

There exists a static solution for this system with a metric
determined by the line element
\beq
ds^2~=~-\left ( 1+c|y| \right)^2d\tau^2~+~dx_idx^i~+~dy^2\ ,
\label{staticform}
\eeq
where the total energy momentum on the brane itself is given by ${\rm
diag} (0,p_b, p_b, p_b)$ and $c = p_b/2M^3_{(5)}$.  The value of the
pressure as a function of the cosmological constant, $\lambda$, is
determined by the equation of state of the matter component on the
brane.  The energy density of that matter component is exactly
balanced by the negative energy density of the cosmological constant.
Note that the metric of the full spacetime explicitly breaks
4-dimensional Lorentz invariance, though if one is confined to the
braneworld, the spacetime appears Minkowskian.

Although this static solution may appear to be fine-tuned, we will
show that for {\em generic} initial states with net positive energy
density, the system dynamically asymptotes to this static solution.
The general
time-dependent line element under consideration is of the form
\begin{equation} \label{cosmback}
ds^{2} = -N^{2}(\tau,y) d\tau^{2}
         +A^{2}(\tau,y)\gamma_{ij}dx^{i}dx^{j}
         +B^{2}(\tau,y)dy^{2}\ ,
\label{metric}
\end{equation}
and, the metric components are given by \cite{Deffayet}
\begin{eqnarray}
N &=& 1 -|y| \ddot{a}\left(\dot{a}^2+k \right)^{-1/2} \nonumber \\
A &=& a -|y| \left(\dot{a}^2+k \right)^{1/2}
\label{bulkmet} \\
B  &=& 1\ , \nonumber
\end{eqnarray}
where $k = -1,0$ or $1$ is the intrinsic spatial curvature parameter.
We take the total energy-momentum tensor which includes matter
and the cosmological constant on the brane to be
\begin{equation}
T^\A_{\quad \B}|_{_{\rm brane}}= ~\delta (y)\ {\rm diag}
\left(-\rho_b,p_b,p_b,p_b,0 \right)\ .
\end{equation}
When the matter content on the brane is specified, the induced scale
factor $a \equiv A(\tau,y=0)$ is determined by the Friedmann equations.

\subsection{Friedmann equations without intrinsic curvature}

For simplicity, let us ignore for the moment the intrinsic curvature
term Eq.~(\ref{branac}) in the action.  The Friedmann equations
derived in
\cite{Kraus:1999it,Binetruy:2000ut,Shiromizu:2000wj,Flanagan:2000cu,bdel} read
\begin{eqnarray}
H^2+\frac{k}{a^2} &=& {1 \over 36 M^6_{(5)}}\rho_b^2
\label{friedfried} \\
\frac{\ddot{a}}{a} &=& 
-{1 \over 36 M^6_{(5)}}\rho_b \left(2\rho_b + 3 p_b\right)\ ,
 \label{ddot}
\end{eqnarray}
where $H$, the Hubble parameter on the brane, is defined by the usual
expression $H ={\dot a\over a}$.  Energy-momentum conservation leads
to \cite{Deffayet,Binetruy:2000ut}
\beq
\dot \rho_b+3{\dot a\over a}(\rho_b+p_b)=0\ .
\label{cons}
\eeq
Using Eqs.~(\ref{friedfried}--\ref{ddot}), the metric components in
Eqs.~(\ref{bulkmet}) can be written as 
\begin{eqnarray}
N &=& 1 + {1 \over 6 M^3_{(5)}} |y| \left(2 \rho_b + 3 p_b \right) \nonumber \\
A &=& a\left(1 - {1 \over 6 M^3_{(5)}} |y| \rho_b \right)
\label{metricvsrho}  \\
B  &=& 1\ . \nonumber
\end{eqnarray}
We may now specify the matter content of our model.
Consider a spatially flat brane ($k=0$), and a brane energy-momentum tensor
given by the sum of that of a cosmological constant $\lambda$ and a
matter energy momentum tensor, such that
\be \label{content}
\rho_b = \lambda + \rho\ ,\ \ \ \ \ 
p_b = -\lambda + p\ ,
\ee
where $\rho$ is the energy density of matter.
We assume that matter obeys the usual equation of state of the form
$p = w \rho$.  In this case the brane energy density is given by
\begin{equation}
\rho_b = \frac{\rho_0}{a^q}+ \lambda\ ,
\end{equation}
where $q = 3(1+w)$ and $\rho_0$ is the initial state matter density.
Implicitly, we have chosen $a(t=t_0) = 1$.  The Friedmann equations
Eqs.~(\ref{friedfried}--\ref{ddot})
demand that the scale factor (for $\lambda \neq 0$) takes the form
\begin{equation}
\rho_0 + \lambda a^q = \left(\rho_0+ \lambda\right)
\exp \left[ \lambda q (t-t_0)/(6M^3_{(5)})\right]\ .
\label{exponential}
\end{equation}
Consider the case when the cosmological constant is negative.
Initially, if the energy density is much larger than the magnitude of
the cosmological constant, the evolution proceeds in the
5-dimensional analog of the big bang scenario: the scale factor
increases as a power of time, the energy density decreases as an
inverse-power with respect to time.  However, as the energy density
decreases, it crosses the threshold of order $|\lambda|$.  The
time-dependence of the scale factor changes.  When the net energy
density is much smaller than the scale of the cosmological constant,
the energy density asymptotes exponentially to zero.  Similarly, the
scale factor asymptotes exponentially to a constant value.  The brane
pressure at late time is simply given by $|\lambda|(1+w)$ so that one sees
from Eq.~(\ref{metricvsrho}) that the metric asymptotes to the
solution given in Eq.~(\ref{staticform}).  The behavior described is
generic.  It is independent of the initial state, so long as the net
energy density of that state is positive.

\subsection{Friedmann equations including intrinsic curvature}

The dynamical behavior just described continues to holds true even in
the presence of an intrinsic curvature term on the brane.  This term is
needed to recover 4-dimensional gravity and cosmology on the brane.
The equations governing the dynamics are modified simply
by replacing the brane energy density $\rho_b$ (respectively pressure
$p_b$) by the sum of the brane energy density $\rho_b$ (pressure
$p_b$) and the effective energy density $\rho_{curv}$ (effective
pressure $p_{curv}$) due to the presence of the brane intrinsic
curvature term.  The quantities $\rho_{curv}$ and $p_{curv}$ are given
by \cite{Deffayet}
\begin{eqnarray}
\label{rhovurv} \rho_{curv}&=& -3 M^2_{(4)}\left\{\frac{\da^2}{a^2}+
\frac{k}{a^2}\right\}  \label{rhocurv} \\ \label{pcurv} p_{curv} &=&
M^2_{(4)} \left\{
\frac{\dot{a}^2}{a^2}   + 2 \frac{\dda}{a}  + \frac{k}{a^2}
\right\}\ . \end{eqnarray}
For example Eq.~(\ref{friedfried}) now reads
\cite{Deffayet} \beq
\sqrt{{H^2} + {k \over a^2}} =  \frac{1}{6M^3_{(5)}}\rho_b - r_0
\left(H^2+\frac{k}{a^2}\right)\ ,
\eeq
which can, in turn, be rewritten
\beq \label{newfried}
\sqrt{H^2+\frac{k}{a^2}} = \frac{1}{2r_0}\left[ -1 +
\sqrt{1 + \frac{2r_0}{3 M^3_{(5)}}\rho_b}\ \right]\ .
\label{hubble_mod}
\eeq
The effect of including the intrinsic curvature term is only felt when the
Hubble parameter is larger than some critical threshold determined by
the parameter $r_0^{-1}$.  Then the cosmological behavior is
4-dimensional.  Eventually, the Hubble parameter decreases below this
threshold, and the system behaves as though the intrinsic curvature
term were not present.  The system inevitably asymptotes to the static
solutions under consideration. This can be seen solving explicitly
the brane Friedmann equation, Eq.~(\ref{newfried}).

Assuming again that the brane content is given by
Eqs.~(\ref{content}) and taking $k=0$, one can solve explicitly for the
scale factor.  Let us define some useful parameters:
\begin{eqnarray}
B &=& 1 - {4\over 3}r_0^2\mu^2	\\
x &=& 1 + {4\over 3}r_0^2\mu^2\left({\rho_b \over |\lambda|}\right)\ ,
\end{eqnarray}
where we have used the mass scale $\mu$ defined in the last section.
Three distinct cases emerge.  If $B < 0$, \beq {q\over 2r_0}
(1-B)(t-t_0) =\ln \left|\frac{x-B}{x-1}
  \frac{1+\sqrt{x}}{1-\sqrt{x}}\right| - 2 \sqrt{-B} \tan^{-1}
\left(\frac{\sqrt{x}}{\sqrt{-B}}\right)\ .  \eeq If $B>0$ and $B\neq
1$, \beq {q\over 2r_0} (1-B)(t-t_0) = \ln \left|\frac{x-B}{x-1}
  \frac{1+\sqrt{x}}{1-\sqrt{x}} \left(
    \frac{\sqrt{B}-\sqrt{x}}{\sqrt{B}+\sqrt{x}}\right)^{\sqrt{B}}\right|\ 
.  \eeq Finally, if $B=1$, \beq {q\over 2r_0}(t-t_0) =
-\frac{\sqrt{x}}{1-x} +\frac{1}{2}\ln
\left|\frac{1+\sqrt{x}}{1-\sqrt{x}}\right| + \frac{1}{x-1}\ .  \eeq
When the brane cosmological constant is negative, one can check that
one recovers the asymptotic behavior mentioned previously. In this
particular case one has $B <1$, and $B\neq 0$, the late time behavior
is then given by considering the limit $x \rightarrow 1$ in the above
equations. On the other hand, the early time limiting behavior is
given by $x \rightarrow \infty$ and can easily be seen to match with
standard cosmology.

The three cases correspond to when the energy scale $\mu$ is larger
than the crossover energy scale ($r_0^{-1}$), when energy scale of the
cosmological constant is smaller than the crossover scale, and when
the cosmological constant is zero, respectively.  If the crossover
energy scale $r_0^{-1}$ is much smaller than $\mu$, we see that a
system with very large initial positive energy density evolves with a
conventional power-law behavior.  As the energy density decreases past
the scale of the cosmological constant, the system evolves as if to
bounce and recollapse in finite time (the acceleration approaches a
constant).  This behavior would occur if the Hubble parameter were
dominated by the second term under the square root in
Eq.~(\ref{hubble_mod}).  However, before that bounce can happen, the
energy density reaches the scale $r_0^{-1}M^3_{(5)}$, the Hubble
parameter depends linearly on the energy density, and the
system asymptotically approaches the static solution, and the bounce
is averted.

\subsection{Scalar field evolution}

\be
     V(\phi) = {1\over 2}m^2\left(\phi^2 - \phi^2_0\right)\ .
\ee
\begin{figure}
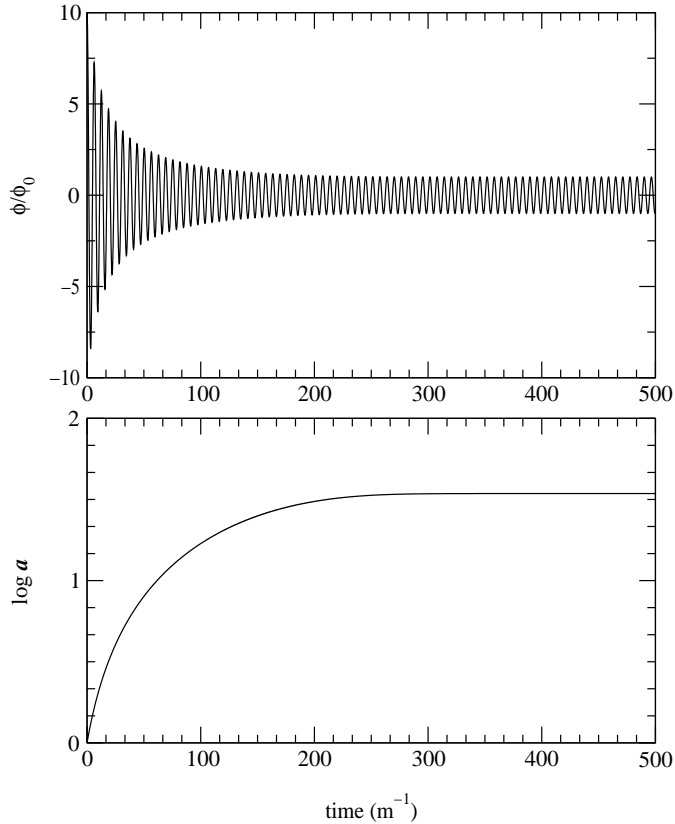
 \begin{center}\PSbox{field.eps
hscale=50 vscale=50 hoffset=-30 voffset=20}{4in}{4in}\end{center}
\caption{
Evolution of the scalar field amplitude and
scale factor for $mr_0 = 1000$ and $r^2_0\mu^2 = 50$.
}
\label{fig:field}
\end{figure}
\begin{figure}
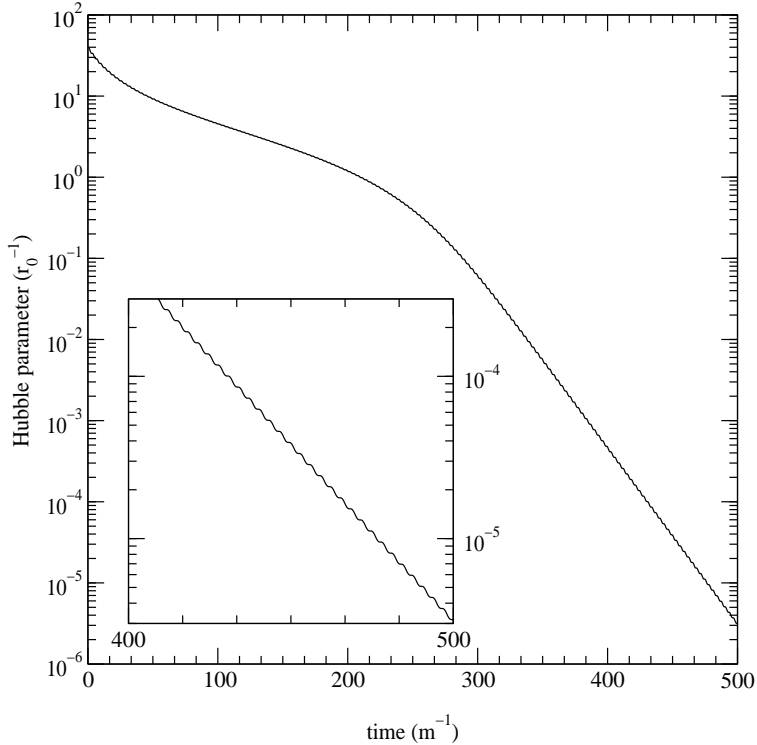
 \begin{center}\PSbox{h.eps
hscale=50 vscale=50 hoffset=-10 voffset=-25}{4in}{3.5in}\end{center}
\caption{
Evolution of the Hubble parameter for the
example shown in Figure 1.  The inset highlights the
details in the evolution that result from the oscillating
field.
}
\label{fig:h}
\end{figure}

This asymptotic dynamics is generic for systems with a negative
cosmological constant and matter on a braneworld.  As another example
of a system that has these elements, consider a scalar field, $\phi$,
living exclusively on the brane with the potential
One can envision that this scalar field may be an inflaton,
quintessence or other such cosmological field.  The constant offset
acts as a negative cosmological constant, with $\lambda =
-m^2\phi_0^2/2$.  Fluctuations in the field around the true vacuum
state act as the matter.  A more general potential may always be
considered, but this example captures the qualitative features of
interest.  Consider the brane homogeneously filled with a scalar field
possessing such a potential.  The equations describing the scalar
field evolutions is
\be
     \ddot{\phi} + 3H\dot\phi + m^2\phi = 0\ ,
\ee
and the Hubble parameter is determined by Eq.~(\ref{newfried}) when
\be
     \rho_b = {1\over 2}\dot\phi^2 + V(\phi)\ .
\ee
Once again, the precise evolution of the system is dependent on two
parameters:  the crossover scale, $r_0$, and the quantity
$r^2_0\mu^2$, and the qualitative features are the same.
Figures~\ref{fig:field} and \ref{fig:h} depict the evolution
of the scalar field, the scale factor, and the Hubble parameter of
a typical case.  With a positive energy density, the field oscillates
with a damping proportional to the Hubble parameter.  Eventually,
the energy density is drawn toward zero, and the scalar field
exponentially asymptotes to a dissipationless oscillation.

One expects that the oscillations in the scalar field amplify
spatially inhomogeneous perturbations in that field.  However, if
the self-coupling of that field is small, such an amplification does
not grow without bound.  Also, unlike the matter considered earlier,
the asymptotic metric of the spacetime is not static.  Although the
energy density vanishes, the pressure is a rapidly oscillating
function.

\section{Global structure of spacetime}

As shown explicitly in \cite{Deffayet,Deruelle:2000ge} (see
also\cite{Mukohyama:2000wi}) the bulk spacetime in Eq.~(\ref{bulkmet})
is two identical pieces of 5-dimensional Minkowski spacetime
glued across the braneworld worldvolume.  The same holds true
for the asymptotic solution, Eq.~(\ref{staticform}), where the bulk is now a
piece of Rindler spacetime.  In this latter case the $y= {\rm constant}$
surfaces in the $(\tau, y)$ plane are hyperbolas ${\cal H}_y$ in the
two dimensional Minkowski space time, so that the brane space-time
(\ref{staticform}) is the hyperbolic cylinder $\cal C$ defined by
${\cal H}_0 \times E_3$, where $E_3$ is the 3 dimensional Euclidian
plane.  The whole spacetime reflected in Eq.~(\ref{staticform}) is then
simply given by gluing two copies of one side of ${\cal C}$ along
${\cal C}$.

This picture of the global structure of spacetime may be verified
explicitly using a new set of coordinates
$Y^A$ ($A=0,1,2,3,5$) of 5-dimensional Minkowski spacetime.  The
line element may be written
\begin{equation} \label{canon}
ds^2 = -(dY^0)^2 + (dY^1)^2 + (dY^2)^2 + (dY^3)^2 + (dY^5)^2\ ,
\end{equation}
where the coordinate change need to arrive at this line element
is determined by equations given in \cite{Deffayet,Deruelle:2000ge}.

\begin{figure} \begin{center}\PSbox{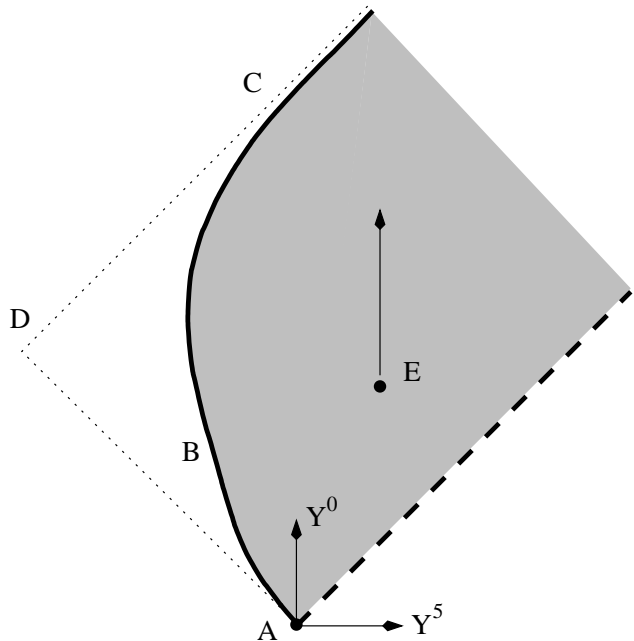
hoffset=0 voffset=0}{3in}{3in}\end{center}
\caption{
Global Minkowski space for a specific spatial slice.  The shaded
region represents the bulk spacetime and the solid line represents
the boundary braneworld worldline on which two identical bulks are
glued together.  (A) For conventional matter content, the braneworld
originates at a big bang singularity which evolves in a lightlike
direction.  (B) The
evolution of the braneworld depends on the specific matter content.
(C) Eventually, the braneworld asymptotes to a future Rindler horizon
and to the static solution.  Equal braneworld
time curves approximately radiate outwards from the light cone origin
at point (D).  Note that a bulk observer (E) is always attracted
to the braneworld and eventually encounters it.
This picture is consistent with the braneworld
point of view that suggests that bulk observers are attracted
to the braneworld from a warp factor that increases as one goes out
into the bulk.
}
\label{fig:global}
\end{figure}

We are mainly interested in the coordinate change for late times
when the braneworld system approaches a static state.  In this
case, whether one considers an intrinsic curvature term Eq.~(\ref{branac})
or not, we
may use the solution in Eq.~(\ref{exponential}) for the scale factor. 
Equation (\ref{exponential}) may be rewritten as
\beq
a = a_\infty(1- \epsilon)^{1/q}\ ,
\eeq 
with $a_\infty$ a constant given by 
\beq
a_\infty = \left(-\frac{\rho_0}{\lambda}\right)^{1/q}
\eeq
is the $t \rightarrow\infty$ limit of the scale factor if $a(t=t_0) = 1$.
The quantity $\epsilon$ defined by
\beq
\epsilon \equiv    \left(1+\frac{\lambda}{\rho_0} \right)e^{-(t-t_0)/T}\ ,
\eeq
which is small for asymptotically late times, and where the time scale $T$ 
 is given by 
\beq
T \equiv -6 M_{(5)}^3 / \lambda q\ .
\eeq   
Using the formulas given in \cite{Deruelle:2000ge}, one finds at leading order
in $\epsilon$ 
\begin{eqnarray}
Y^0+Y^5 - \frac{a_\infty r^2}{2} &=& \epsilon^{-1} \frac{q T^2}{2 a_\infty}
	\left(1+ \frac{|y|}{T} \right), \\
Y^0-Y^5 -2 a_\infty &=& - \epsilon  \frac{2a_\infty}{q} 
\left(1+ \frac{|y|}{T} \right), \\
Y^i &=& a_\infty x^i\ , 
\end{eqnarray}
where $r = x^i x_i$.
One then easily sees that up to terms of order $\epsilon$ (and a change
in the origin of the Y-coordinate system), one has for a given $r = {\rm
constant}$ slice
\beq
{(Y^0)}^2 - {(Y^5)}^2 = - T^2 \left(1+\frac{|y|}{T}\right)^2\ .
\eeq
For asymptotically late times, the world line of the braneworld
follows this hyperbola.  At specific time slices, the braneworld
coordinates, $Y^0$ and $Y^5$ evolve toward infinity asymptoting
the light cone, and the values of those coordinates grow
exponentially with cosmological (braneworld) time $t$.
Figure~\ref{fig:global} depicts the global 5-dimensional
Minkowski space for a given spatial slice.

\section{Four-dimensional gravity}

\subsection{Propagator on the brane}

We can now study the propagation of gravitons in this asymptotic static
space.  The line element reads as follows: 
\beq
ds^2~=~-\left ( 1+r_0\mu^2|y|  \right )^2d\tau^2~+~dx_idx^i~+~dy^2~\ .
\label{sol1}
\eeq
For simplicity we treat the graviton as a scalar particle.  As
was emphasized in \cite{Dvali:2000hr,Bajc:2000mh}, the scalar part of the
graviton propagator reduces to that of a minimally coupled spin-0
particle in various backgrounds which we study below.

Let us start with the action:
\beq
\int ~d^4x~dy \left ( ~{\5M \over 2} \sqrt {{\tilde g}} ~{\tilde g}^{AB}
\partial_A \phi~ \partial_B \phi~+~
{\4M \over 2} ~\delta(y) ~\sqrt {{g}}~{g}^{ab} 
\partial_a \phi~ \partial_b \phi~\right )~.
\label{model}
\eeq
The first term here is a counterpart of the bulk Einstein action  
and the second one accounts for the induced curvature term.
The equation for the Green function takes the form:
\beq
\left ( {\5M}~\partial_A \sqrt {{\tilde g}}
\partial^A  ~+~\delta(y) {\4M}~  \partial_b \sqrt {{g}}  
\partial^b \right ) G(x_a,y)~=~\5M \sqrt {g}\delta^{(4)}(x)\delta(y)~.
\eeq
Using (\ref {sol1}) we find:
\bea
\left \{ -{1\over ( 1+r_0\mu^2|y|)}\partial_\tau^2
~+~ ( 1+r_0\mu^2|y|)\partial_i^2~+~
\partial_y( 1+r_0\mu^2|y|)\partial_y \right \}~G(x_a,y)~ &+&\nonumber \\
2r_0 \delta(y) \left \{  -{1\over ( 1+r_0\mu^2|y|)}\partial_\tau^2~+~ 
( 1+r_0\mu^2|y|)\partial_i^2~ \right \}G(x_a,y) &=&\delta^{(4)}(x)\delta(y)~.
\label{gr1}
\eea
To find the Green's function it is convenient to
perform   Fourier transform to momentum space 
with respect to the 
four worldvolume coordinates $x_a$,
\beq
G(\tau,x_i,y)~\equiv~\int {d\omega~d{\bf q}_i \over (2\pi)^4}
e^{i\omega\tau}e^{-i{\bf q}\cdot{\bf x}}~{\tilde G}(\omega,{\bf q}_i,y)~.
\label{four}
\eeq
As a result, the equation takes the form:
\bea
\left \{ {\omega^2\over ( 1+r_0\mu^2|y|)}~-~ ( 1+r_0\mu^2|y|){\bf q}_i^2~+~
\partial_y( 1+r_0\mu^2|y|)\partial_y \right \}~{\tilde G}(\omega,q,y) 
&+&\nonumber \\
2r_0 \delta(y) \left \{  \omega^2~-~ 
{\bf q}_i^2~ \right \}{\tilde G}(\omega,q,y) &=&\delta(y)~.
\label{gr2}
\eea
Let us introduce the following notation:
\beq
{\tilde G}~\equiv~{1\over \sqrt{1+r_0\mu^2|y| }}\chi~.
\label{GChi}
\eeq
Making use of this relation,
the resulting equation reads as follows:
\beq
\chi^{\prime\prime}~+~
\left [{\omega^2~+~(r_0\mu^2)^2/4\over (1+r_0\mu^2|y|)^2 }-{\bf q}_i^2 
\right ]~\chi ~+~
2r_0 \delta(y)\left \{  \omega^2~-~ 
{\bf q}_i^2~ \right \}~\chi~=~\delta(y) \left [1+r_0\mu^2\chi(0)\right]~.
\label{chi1}
\eeq
To find the solution of this equation with appropriate boundary conditions
let us introduce the following substitution:
\beq
\chi(\omega,{\bf q}_i,y)~=~B(\omega,{\bf q}_i)~\Sigma (y)~,
\label{ChiSigma}
\eeq
where $\Sigma$ satisfies the 
following equation:
\beq
\Sigma^{\prime\prime}~+
~\left [{\omega^2~+~(r_0\mu^2)^2/4\over (1+r_0\mu^2|y|)^2 } - {\bf q}_i^2
\right ]~\Sigma ~=~\delta(y)~.
\label{Sigma1}
\eeq
From this definition, as well as from Eq.~(\ref{ChiSigma}), one 
sees that the expression for the Green's function on the brane ($y=0$)
takes the form:
\beq
{\tilde G}|_{y=0}~=~
 ~\left[- 2r_0q^2 - r_0\mu^2 + \Sigma(0)^{-1} \right ]^{-1}\ ,
\label{prop1}
\eeq 
where we have defined $q^2 = -\omega^2 + {\bf q}_i^2$,
the invariant 4-momentum squared.

The solution to Eq.~(\ref{Sigma1}) with asymptotically vanishing
behavior as $|y|\rightarrow \infty$ is \cite{Watson}
\beq
\Sigma(y) \sim \sqrt{u}K_\nu(u/2)\ ,
\eeq
with
\beq
u = \frac{1+r_0 \mu^2 |y|}{r_0 \mu^2} 2|{\bf q}_i|\ ,
\eeq
and $K_\nu$ is the MacDonald function of order
$\nu = \pm i\omega/r_0\mu^2$.  The normalization
of $\Sigma(y)$ may be established by the delta function
in the right-hand side of Eq.~(\ref{Sigma1}), indicating
that
\beq
\Sigma^\prime(y = 0^+) = \frac{1}{2}\ .
\eeq
One obtains 
\beq
\Sigma(0)^{-1} = 2 |{\bf q}_i|\left[\frac{1-2\nu}{u_0} - 
\frac{K_{\nu-1}(u_0/2)}{K_\nu(u_0/2)}\right]\ ,
\eeq 
where we define
\beq
u_0 \equiv u(y=0) = \frac{2|{\bf q}_i|}{r_0 \mu^2}\ .
\eeq
Using Eq.~(\ref{prop1}), one finds the exact expression of the
propagator on the brane:
\beq
{\tilde G}|_{y=0}~=~
 ~ \frac{1}{2r_0}\left[-q^2
\mp \frac{i\omega}{r_0} -\frac{|{\bf q}_i|}{r_0}
\frac{K_{ -1 \pm  i \omega / r_0 \mu^2}
\left(|{\bf q}_i| / r_0 \mu^2 \right)}{K_{\pm  i \omega / r_0 \mu^2}
\left(|{\bf q}_i| / r_0 \mu^2\right)}
\right]^{-1}\ .
\label{propprop}
\eeq
In the above expression the first term on the right hand
side represents the usual 4-dimensional Lorentz invariant
propagator, whereas the two other terms are responsible for deviation
from 4-dimensional behavior as well as for Lorentz violating
effects.\footnote{Because the MacDonald function is even with respect
to its order, one can indeed verify that one arrives at the same
result regardless of the sign taken in front of $\omega$ in
Eq.~(\ref{propprop}).}

\subsection{Asymptotic developments}

We wish to elaborate on the propagator Eq.~(\ref{propprop}) in
some special limits.  We see that
the propagator is 4-dimensional Lorentz symmetry violating
in general.  Let us first examine the limit in which
the spatial momentum ${\bf q}_i \rightarrow 0$, with $\omega \ne 0$.
Taking the propagator with outgoing radiative boundary conditions,
at leading order
\be
{\tilde G}|_{y=0}^{-1}~=~ 2r_0\omega^2 - 2i\omega\ .
\label{propomega}
\ee
We see then a restoration of the propagator found in \cite{Dvali:2000hr}
in the same limit, indicating there is no massive graviton state
stationary with respect to the cosmological frame.

Next we wish to take the limit where the propagator yields the
static potential, $\omega \rightarrow 0$.
\be
{\tilde G}|_{y=0}^{-1}~=~ -2r_0{\bf q}_i^2
	- 2|{\bf q}_i
	\frac{K_1(|{\bf q}_i|/r_0\mu^2)}{K_0(|{\bf q}_i|/r_0\mu^2)}\ .
\ee
Note that this form is not covariant with respect to the form found
in Eq.~(\ref{propomega}).  For large distances
($|{\bf q}_i| \rightarrow 0$), the propagator reduces to
\be
{\tilde G}|_{y=0}~=~ \frac{1}{2r_0\mu^2}
	\ln\frac{|{\bf q}_i|}{2r_0\mu^2}\ .
\ee
This form corresponds to a
$r^{-3}$ static potential.  For short distances,
($|{\bf q}_i| \rightarrow \infty$),
the propagator reduces to
\be
{\tilde G}|_{y=0}~=~ - \frac{1}{2r_0{\bf q}_i^2}\ ,
\ee
corresponding to a $r^{-1}$ static potential.  Crossover behavior
varies depending on the value of the dimensionless parameter, $r_0\mu$.

In order to see the restoration of 4-dimensional Lorentz invariance,
we must go to large values of $\omega$ and ${\bf q}_i$.  Taking $r_0\mu^2
\rightarrow 0$ while holding $\omega$ and ${\bf q}_i$ fixed, the
propagator becomes
\be
{\tilde G}|_{y=0}^{-1}~=~
 - 2r_0q^2  - 2 q - r_0\mu^2(1 + \omega^2/q^2)\ .
\label{limit1}
\ee
Note that when $r_0\mu^2\rightarrow 0$, explicit 4-dimensional
Lorentz invariance is recovered and the propagator reduces to the form
found in \cite{Dvali:2000hr}.  Equation~(\ref{limit1}) expresses the
dominant off-shell behavior for the scalar graviton propagator.

Alternatively, when $q^2 \rightarrow 0$, but in a manner such that
$\omega^2 \gg r_0^2\mu^4$ with $q/\omega < (r_0\mu^2/\omega)^{1/3}$, then
\be
{\tilde G}|_{y=0}^{-1}~=~ - 2r_0q^2
-2 \omega~\frac{\Gamma(2/3)}{\Gamma(1/3)}
\left({6 r_0\mu^2 \over \omega}\right)^{1/3}
\ .
\ee
The pole in the propagator in this regime indicates the following
dispersion relationship for an on-shell particle:
\be
\omega^2 = {\bf q}_i^2 + \frac{\Gamma(2/3)}{\Gamma(1/3)}
\left(6 r_0\mu^2\right)^{1/3}({\bf q}_i^2)^{1/3}\ .
\ee
As $|{\bf q}_i| \rightarrow
\infty$, the first term in the expression is dominant and the
expected relativistic dispersion relationship is recovered.

\section{Concluding Remarks}

Our system has the property of having energy density evolve to
zero from a generic set of initial conditions with positive energy
density.  The pressure of this asymptotic state is non-zero, so
the full 5-dimensional spacetime explicitly breaks 4-dimensional
Lorentz invariance, even while the induced metric on the braneworld
is Minkowskian.  The graviton propagator on such
a braneworld also reflects a lack of 4-dimensional Lorentz 
symmetry, since that graviton can probe the full spacetime.

A static Minkowski braneworld coupled
to a braneworld energy-momentum which explicitly violates
4-dimensional Lorentz symmetry is possible because the larger metric
of the 5-dimensional space does not respect 4-dimensional Lorentz
symmetry.  Any slice of the 5-dimensional spacetime parallel to the
braneworld is indeed Minkowski.  However, the speed of light varies
from slice to slice.  If standard model particles trapped on the
braneworld do not couple to the dark matter generating the
5-dimensional spacetime, then there should be no observable Lorentz
symmetry violation in the standard model, except through graviton
exchange.

One can ask how well this model complies with known phenomenology.
We have two parameters that may be adjusted: the crossover scale, $r_0
= M^2_{(4)}/2M^3_{(5)}$, which reflects the relative sizes of the
4-dimensional Planck mass and the 5-dimensional Planck mass, as
well as $\mu^2 = -\lambda/M^2_{(4)}$, which reflects the relative size
of the brane cosmological constant to the 4-dimensional Planck
scale.

We may place constraints on the system parameters at various stages.
Let us first address solely the constraints from gravitational force
law observations.  The exchange of the tower of Kaluza-Klein (KK)
modes may be understood from the 4-dimensional point of view as
an exchange of a metastable graviton \cite{quasi1,CEH,Dvali:2000rv}.
The decay width is controlled by the larger of the two energy scales
$r_0^{-1}$ or $\mu$.  Thus, both energy scales must be tuned to some
sufficiently small value to be consistent with a stable, massless
graviton.  We know that the graviton is stable and massless up until
$\sim 100$~parsecs (for a discussion, see \cite{Will:1998bb}).
We may choose the galactic size as well, but for
here we settle for a strict confirmation of the $1/r^2$ force law.  Then
both $r_0$, and $\mu^{-1}$ must be as large as $100$~parsecs.
This places constraints on $M_{(5)} < 100\ {\rm GeV}$ and $|\lambda| <
(10\ {\rm eV})^4$.

We may also constrain our system parameters based on the standard
cosmological model.  We found that early cosmology only follows the
conventional hot big bang scenario when the energy density on the
brane is much larger than $|\lambda|$ as well as a similar constraint
between the Hubble parameter and the crossover scale $r_0^{-1}$.
Cosmology constraints imply $M_{(5)} < 100\ {\rm MeV}$ and $|\lambda| <
(10^{-3}\ {\rm eV})^4$.

Having such a small fundamental Planck scale, $M_{(5)}$, does not
contradict any astrophysical or particle physics bounds as long as
there is a 4-dimensional induced Planck scale on the brane (see
\cite{Dvali:2001gm}), assuming quantum gravity becomes soft above this
energy scale.  However, in the present context this issue deserves
clarification, due to the presence of a Lorentz symmetry violating
metric which is not considered in \cite{Dvali:2001gm}.  Since the Lorentz
violation is due to a tiny cosmological constant, the prior analysis
should be valid, and any new effect should be a small perturbation.
On the other hand, the warp factor (which parametrizes Lorentz
symmetry violation) diverges as $M_{(5)} \rightarrow 0$.  While this
statement is true, Lorentz symmetry violation in the propagator on the
brane is controlled by the second and thirds terms on the right-hand
side of Eq.~(\ref{propprop}).  For small $M_{(5)}$ (when $M_{(4)}$ and
$\lambda$ are fixed) those terms are suppressed relative to the
leading 4-dimensional contribution to the propagator.  In other words,
although the Lorentz violating warp factor becomes steep when
$M_{(5)}$ is small, it cannot strongly affect brane observers since
probing the bulk becomes more difficult.  This can be also understood
in the KK-mode expansion.  In this language, the intrinsic curvature
term repels KK-modes heavier than $1/r_0$, suppressing their
wavefunction on the brane and shielding the brane from bulk gravity
effects, in particular Lorentz symmetry violation.

All these observations imply that one cannot start with a large
cosmological constant on the brane, and therefore this model cannot
yield a resolution of the cosmological constant problem.  The
mechanism that so generously compels the total energy density to zero
is equally unforgiving.  Any combination of matter whose total energy
density is on the order of magnitude of the cosmological constant
evolves exponentially fast to a flat and perfectly static universe, in
a time comparable to $M^3_{(5)}/|\lambda|$.

We have proposed a model which has some intriguing properties.  It
dynamically evolves to a zero energy-density state and provides a
novel mechanism for producing a static, Lorentz symmetry violating universe.
Perhaps these mechanisms may be applied to more phenomenologically
promising models in the future.

\acknowledgments

We would like to thank C. Grojean and D. Hogg for discussions.  C. D. and
G. G. thank the LBL Theory Group for its hospitality where part of
this work was done.  This work is sponsored in part by NSF Award
PHY-9803174 and PHY-0070787, David and Lucille Packard Foundation
Fellowship 99-1462, the Alfred P. Sloan Foundation Fellowship, and DOE
Grant DE-FG02-94ER408.

\end{document}